# Scalable Fabrication of Thermally Reconfigurable Ge Metasurfaces Using Stencil Lithography for Mid-Infrared Molecular Sensing

Shovasis Kumar Biswas[1,2‡,] Samir Rosas[1,‡], Wihan Adi[1], Aidana Beisenova[1], Vaishakh Unnikrishnan[1,2], Furkan Kuruoglu[1,3], Filiz Yesilkoy[1]*

[1] Department of Biomedical Engineering, University of Wisconsin-Madison Madison, WI 53706, USA

[2] Department of Electrical and Computer Engineering, University of Wisconsin–Madison, Madison, WI 53706, USA

[3] Department of Physics, Faculty of Science, Istanbul University, Vezneciler, 34134, Istanbul, Turkey

*Corresponding author. Email: filiz.yesilkoy@wisc.edu

‡ Authors contributed equally

**Abstract**

Mid-infrared (mid-IR) spectroscopy enables label-free molecular detection and is widely employed in biomedical, environmental, and chemical sensing; however, its broader deployment remains limited by bulky instrumentation and insufficient analytical sensitivity. Photonic metasurfaces supporting strong mid-IR resonances provide a promising route toward compact on-chip spectrometers and enhanced molecular sensing, yet their practical implementation is often constrained by limited spectral coverage and fabrication complexity. Here, we present a scalable, resist-free stencil lithography approach for fabricating arrays of Ge pillar metasurfaces on $CaF_2$ substrates that support polarization-insensitive Mie resonances dominated by electric dipole modes. By varying the geometric design parameters during fabrication, we engineered metasurface arrays with discrete resonances spanning the 950–1700 $cm^{-1}$ molecular fingerprint region. Furthermore, leveraging the thermo-optic response of Ge, we achieved dynamic tuning of these discrete resonances, demonstrating continuous and reversible resonance shifts of approximately 36 $cm^{-1}$ per metasurface over 300–500 K, corresponding to a tuning rate of ~0.18 $cm^{-1}$ $K^{-1}$. The thermally induced spectral sweeping enables multiplexed detection of poly(methyl methacrylate) vibrational modes and continuous reconstruction of its absorbance spectrum across 1100–1215 $cm^{-1}$. These results establish a scalable dielectric metasurface platform with spectrally reconfigurable mid-IR modes for molecular sensing across the fingerprint region and compact infrared sensor technologies.

## Introduction

Light–matter interactions in the mid-infrared (mid-IR) are inherently frequency dependent because this spectral region overlaps with fundamental molecular vibrational modes. As a result, mid-IR spectroscopy offers a powerful label-free approach for molecular analysis; however, its analytical sensitivity is fundamentally constrained by the small absorption cross-sections of most molecular species[1,2]. A viable strategy to overcome this limitation is to enhance local electromagnetic fields using photonic cavities formed by metallic antennas[3–6], dielectric resonators[7,8], or their ordered two-dimensional arrays (metasurfaces)[9]. Thus, surface-enhanced infrared absorption spectroscopy (SEIRAS) has emerged as a powerful platform for improving both the sensitivity and selectivity of conventional mid-IR vibrational spectroscopy[10–13].

SEIRAS has been most commonly implemented using plasmonic metasurfaces because of their strong field confinement and broadband resonances[14–17]. Although these characteristics enable broad spectral coverage, the spectrally non-uniform field enhancement associated with broad resonances leads to uneven amplification of molecular fingerprints. This can be tolerated when quantifying known analytes, but it becomes a significant limitation for interrogating chemically complex samples with unknown spectral features. Dielectric metasurfaces composed of low-loss resonators, such as silicon (Si) or germanium (Ge), provide an attractive alternative by supporting high-quality-factor resonances together with strong electromagnetic confinement[8]. These low-loss metasurfaces have enabled highly sensitive retrieval of molecular fingerprints, uniformly enhancing a broad vibrational spectrum with a comb of narrow high-Q resonances. However, generating a dense array of sharp resonances that sweep the fingerprint spectrum typically requires large area metasurface arrays[8] or gradient architectures[18,19] with spatially multiplexed spectrally engineered resonators. Consequently, scalable fabrication of dielectric metasurfaces over large areas remains a major bottleneck for the broader deployment and technological translation of SEIRAS platforms.

To address the limited throughput of conventional dielectric metasurface fabrication, deep ultraviolet lithography (DUVL) has been proposed as a scalable manufacturing strategy[20]. Nevertheless, established DUVL workflows are predominantly tailored to silicon-based wafers, which impose material and processing constraints for the realization of mid-IR resonator arrays[21]. In particular, optimal resonant performance in dielectric metasurfaces is typically achieved when resonators are fabricated on IR-transparent solid substrates, such as Calcium fluoride or Barium fluoride; however, achieving scalable fabrication on these platforms remains challenging. Consequently, alternative fabrication strategies, including nanoimprint lithography and stencil-based shadow-mask evaporation, have been proposed[22–25]. Among these, stencil lithography[26] is especially attractive because it enables direct, single-step deposition of resonator materials through shadow-mask apertures without dependence on substrate-specific process compatibility, thereby providing a rapid and cost-effective route to large-area metasurface fabrication. Yet, its application to SEIRAS platforms remains largely unexplored, despite its strong potential to enable scalable high-index dielectric metasurfaces for biochemical sensing.

A further limitation of most existing metasurfaces is their intrinsically static response, which precludes post-fabrication alignment of resonance frequencies with characteristic molecular vibrational bands, a critical requirement for efficient SEIRAS. Gradient metasurfaces partially address this by spatially distributing resonances through continuous scaling of the metasurface geometry[27,28]. However, they do not allow for post-fabrication dynamic tuning, thereby preventing selective alignment of a metasurface resonance with a specific molecular mode or compensation for spectral shifts arising from fabrication imperfections and environmental fluctuations. Dynamically tunable dielectric metasurfaces offer a potential route to overcome this limitation by enabling continuous spectral matching[29,30]. Electrically tunable mid-IR metasurfaces based on III–V semiconductors[31], graphene[32,33], and van der Waals materials[34] have previously been demonstrated; however, their practical implementation remains constrained by limited scalability and broadened resonances that compromise spectral selectivity. Likewise, reconfigurable metasurfaces based on phase-change materials[35,36] are poorly suited for SEIRAS because the localized field must overlap with the phase-change media rather than the target analyte. These limitations underscore the need for a new class of reconfigurable metasurfaces that combines post-fabrication tunability across the fingerprint spectrum with low-cost, scalable fabrication strategies, thereby enabling next-generation biochemical sensing and mid-infrared photonic technologies.

In this work, we demonstrate SEIRAS using dynamically tunable dielectric metasurfaces fabricated over large areas by scalable stencil lithography (Figure 1). Ge pillar arrays were deposited on $CaF_2$ substrates through engineered stencil apertures. The masks were mass manufactured on Si substrates by patterning silicon nitride membranes using standard UV lithography. Arranged in a hexagonal lattice, the Ge pillars support polarization-insensitive resonances governed by electric-dipole-dominated Mie-type modes. In a single fabrication step, we fabricated arrays of metasurfaces with discrete resonances spanning the 950–1700 $cm^{-1}$ molecular fingerprint region. To extend this discrete spectral coverage into continuous tunability, the resonances were thermally shifted by exploiting the thermo-optic response of Ge. Over a temperature range of 300–500 K, each metasurface exhibited a continuous and reversible resonance shift of ~36 $cm^{-1}$ (tuning rate ~0.18 $cm^{-1}$ $K^{-1}$), enabling active spectral control without auxiliary tuning materials or complex device architectures. This thermal sweeping strategy enabled multiplexed detection of Poly(methyl methacrylate) vibrational modes at 1148 and 1192 $cm^{-1}$ and continuous reconstruction of its absorbance spectrum across 1100–1215 $cm^{-1}$. By combining scalable fabrication with dynamically tunable sharp metasurface resonances across the fingerprint region, this platform provides a practical route toward large-area, reconfigurable metasurface sensors for mid-IR spectroscopy, with potential applications in environmental monitoring, biomedical diagnostics, and chemical analysis.

## RESULTS AND DISCUSSION

### *Scalable fabrication of gradient Ge pillar metasurfaces using stencil lithography*

The fabrication workflow is illustrated in Fig. 1a–c. Ge pillar array metasurfaces were fabricated by depositing a 600 nm Ge layer onto calcium fluoride substrates through a stencil mask using highly directional electron-beam evaporation. The stencil mask was tightly clamped onto the substrate to minimize mask–substrate separation during deposition and ensure accurate pattern transfer. Following deposition, the mask was mechanically removed without chemical solvents or lift-off processing, leaving behind ordered arrays of Ge resonators directly patterned on the substrate.

Importantly, the stencil masks were fabricated by high-throughput UV lithography on 6-inch wafers, enabling parallel and low-cost production of multiple masks. Each mask consists of aperture arrays patterned in a 400 nm freestanding silicon nitride membrane supported by a silicon frame. By varying aperture diameter and lattice periodicity across the mask, multiple metasurface geometries were defined within a single chip (Fig. 1d). Each region was designed with a unique geometry to generate resonances spanning the mid-IR fingerprint region from 950 to 1700 cm$^{-1}$. The large-area uniformity of the fabricated structures is evident in the optical micrograph shown in Fig. 1e, while the scanning electron micrograph in Fig. 1f reveals uniform Ge pillars with well-defined edges and minimal structural defects.

To evaluate the optical response, reflectance spectra were measured under normal IR incidence for metasurfaces $\mathrm{MS}_1 - \mathrm{MS}_8$ (Fig. 1g). Each metasurface exhibits a distinct resonance whose spectral position shifts progressively from ~1700 cm$^{-1}$ ($\mathrm{MS}_1$) to ~1000 cm$^{-1}$ ($\mathrm{MS}_8$) as a function of lateral geometry scaling. The narrow linewidths and well-defined spectral peaks indicate strong optical confinement and low radiative loss, consistent with electric-dipole-dominated Mie-type resonances in high-index dielectric resonators[37].

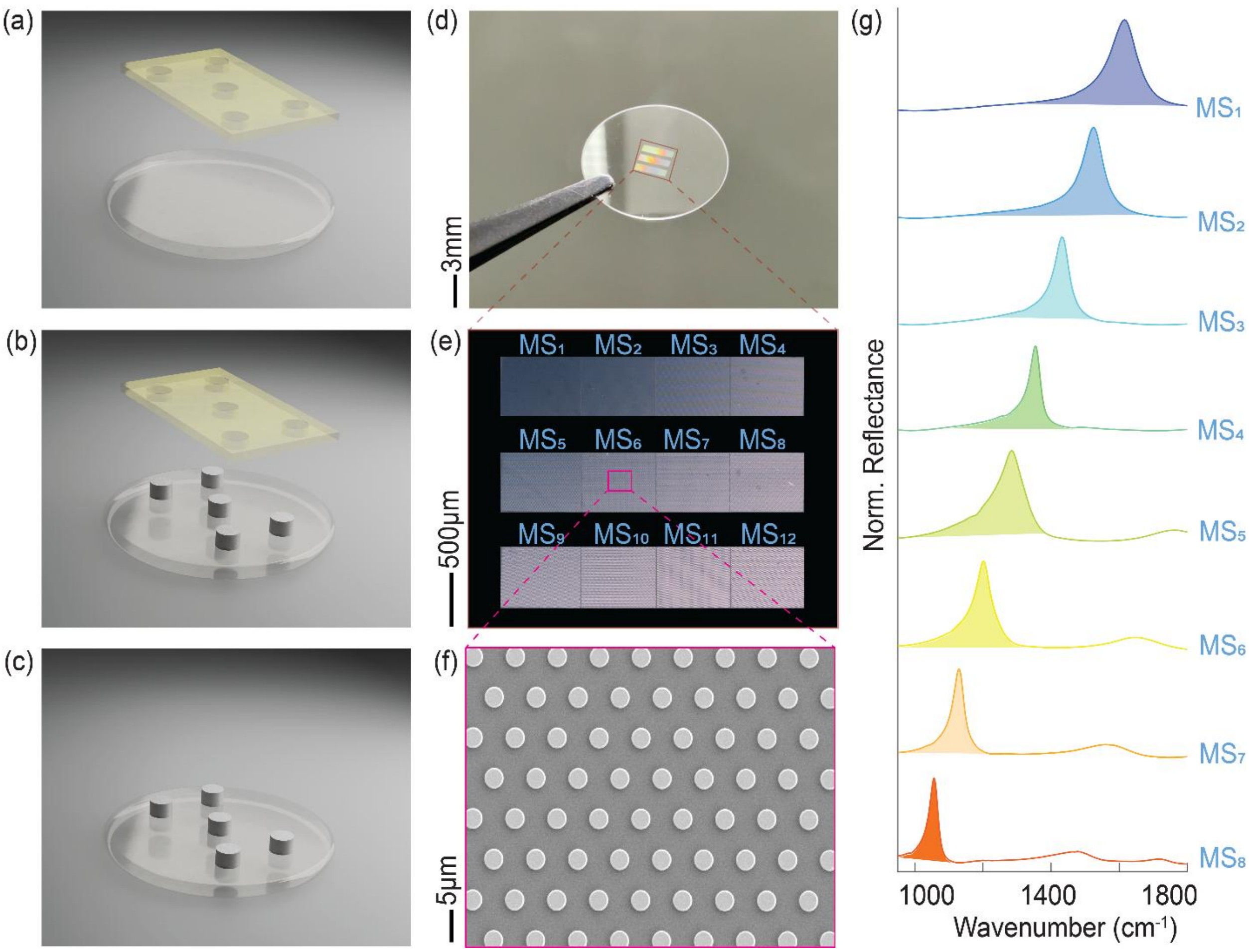

**Figure 1. Large area, scalable fabrication of Ge pillar metasurfaces using stencil lithography.** a–**c,** Schematics showing step-by-step fabrication process. A stencil mask with periodic circular

apertures is clamped over a $CaF_2$ substrate **(a)**, followed by a directional deposition of Ge through the mask to define vertical hexagonal pillar arrays **(b)**. After mask removal, a periodic Ge pillar metasurface remains on the substrate **(c)**. **d, e** Optical images of the fabricated metasurface array, showing 12 different hexagonal pillar array designs ($\mathrm{MS}_1 - \mathrm{MS}_{12}$), each exhibiting a distinct resonance in the mid-infrared, spanning ~950 $cm^{-1}$ to 1700 $cm^{-1}$. **f,** scanning electron microscope (SEM) image of the fabricated metasurface, showing a uniform hexagonal array of Ge pillars with high structural fidelity. **g,** Normalized measured reflectance spectra for metasurfaces, $\mathrm{MS}_1 - \mathrm{MS}_8$, illustrating resonance peaks tuned across a broad spectral range from ~1700 $cm^{-1}$ to ~1000 $cm^{-1}$.

***Photonic resonance properties of the Ge pillar array metasurfaces***

To investigate the physical mechanism underlying the Ge pillar array metasurfaces, we performed electromagnetic simulations and multipolar mode decomposition analysis (Fig 2). A major strength of our design lies in its geometric scalability, which enables spectral tuning of the resonance across the fingerprint spectrum while preserving the resonance properties. To demonstrate resonance control via geometric tuning, we introduced a scaling factor, S, which modifies the lateral dimensions of the unit cell, including lattice period (Px, Py) and pillar radius (r), while maintaining a constant pillar height of h = 600 nm (Fig. 2a). This approach preserves the overall mode properties while shifting the resonance frequency.

The simulated and experimentally measured reflectance spectra for the geometrically scaled metasurfaces are presented in Fig. 2c,d. Excellent agreement is observed between simulation and experiment in both resonance position and linewidth across the scaling sequence. A systematic redshift of the resonance from higher to lower wavenumbers is obtained with increasing lateral dimensions, confirming that spectral tuning is governed primarily by geometric scaling. Raw reflectance spectra (Supplementary Fig. S1) further show peak reflectance amplitudes exceeding 70% for all metasurfaces, indicating strong far-field coupling and high optical quality of the fabricated structures. Importantly, the resonance linewidth remains narrow throughout the tuning range, demonstrating that spectral tuning does not significantly degrade modal confinement.

To identify the modal origin of the resonance, multipole decomposition of the scattered field was performed for a representative metasurface design (Fig. 2e). The analysis reveals that the resonance is dominated by the electric dipole (ED) contribution, with only minor contributions from magnetic dipole (MD), electric quadrupole (EQ), and magnetic quadrupole (MQ) components. This dipolar character is further confirmed by the simulated electric-field enhancement profile for a metasurface resonating at 1654 $cm^{-1}$ (Fig. 2b). The normalized field amplitude ($|E|/|E_0|$) exhibits strong localization near opposite edges of the Ge pillar along the incident polarization axis, producing well-defined hotspots at the pillar boundary. These localized enhancements define regions of maximal analyte–field interaction and are particularly favorable for SEIRAS.

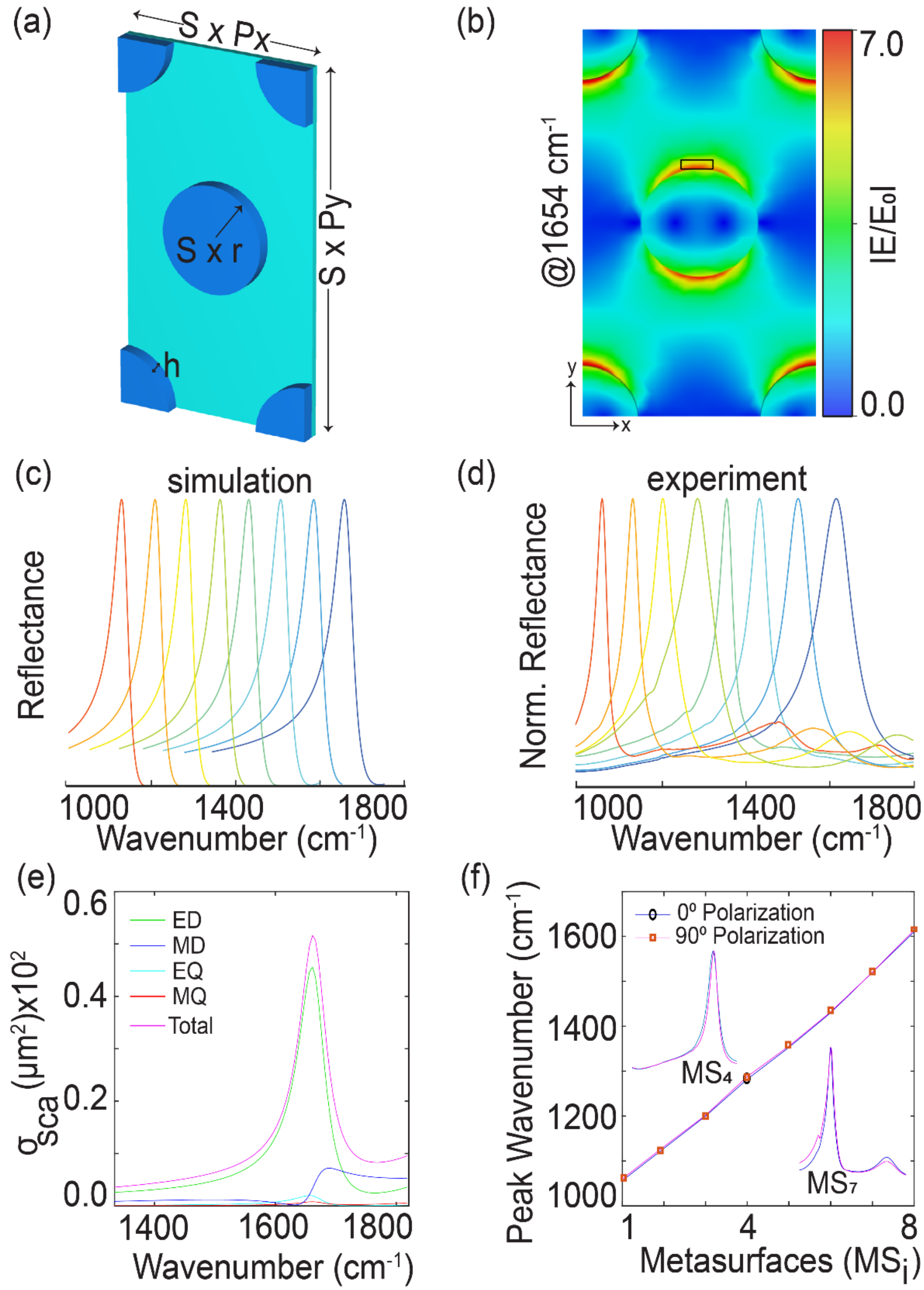

**Figure 2. Photonic resonance characteristics of the Ge pillar metasurface**. **a**, Schematic representation of the metasurface unit cell with periods $P_x$ = 4.722 µm and $P_y$ = 8.44 µm, pillar radius r = 1.154 µm, and pillar height h = 0.6 µm. By varying the geometric scaling factor S from 2.25 to 1.0, we tune the resonance wavenumber across a spectral range of 950–1700 cm$^{-1}$. **b,** Electric field enhancement map, $|E|/|E_0|$, calculated at the resonance wavenumber of 1654 cm$^{-1}$, showing localized field-enhancement around the pillar sidewalls, indicative of an electric

dipole resonance. **c,** Simulated and **d,** experimentally measured reflectance spectra, showing systematically tuned resonance peaks across the mid-IR region from ~1000 to ~1700 $cm^{-1}$. **e,** Multipole decomposition of the simulated scattering cross-section, revealing dominant electric dipole (ED) contribution with minor contributions from magnetic dipole (MD), electric quadrupole (EQ), and magnetic quadrupole (MQ) modes. **f,** Experimental peak positions of resonances as a function of metasurface index ($\mathrm{MS_i}$) for orthogonal polarizations (0° and 90°), confirming linear spectral tunability and polarization-insensitive behavior. Insets show representative normalized reflectance spectra of $\mathrm{MS_4}$and $\mathrm{MS_7}$ under both polarizations.

The polarization response was experimentally assessed by measuring reflectance under orthogonal linear polarizations (0° and 90°). As shown in Fig. 2f, the resonance positions of metasurfaces $\mathrm{MS_1} - \mathrm{MS_8}$ remain nearly unchanged under both polarization states. Insets for $\mathrm{MS_4}$ and $\mathrm{MS_7}$ further confirm identical spectral response with minimal variation in resonance amplitude or linewidth. This polarization robustness arises from the structural symmetry of the pillar geometry and is advantageous for sensing configurations where polarization control is limited.

***Thermally induced dynamic spectral tuning of Ge pillar metasurfaces***

To demonstrate dynamic thermal tunability of Ge pillar array metasurfaces, we investigated eight metasurface regions ($\mathrm{MS_1} - \mathrm{MS_8}$) (Fig. 3a), each featuring a distinct unit cell geometry. The normalized reflectance spectra from $\mathrm{MS_1} - \mathrm{MS_8}$ measured at room temperature are displayed in Fig. 3 b, (top panel). Each metasurface exhibits a spectrally distinct and sharp resonance across a broad spectral window from ~1000 $cm^{-1}$ to ~1700 $cm^{-1}$. To evaluate post-fabrication dynamic tunability, we examined the thermal responsiveness of the same metasurface array by increasing the temperature from 300 K (room temperature, RT) to 500 K by steps of 20 K. As illustrated in Fig. 3b (bottom panel), all metasurfaces exhibited a consistent redshift in resonance frequency with increasing temperature. This shift arises from the intrinsic thermo-optic effect of Ge, whose refractive index increases with temperature, thereby altering the resonance condition[38,39]. Across the full temperature range, each metasurface exhibited a spectral shift of 36 $cm^{-1}$, corresponding to a tuning rate of 0.18 $cm^{-1}/K$. This dynamic spectral control provides an additional degree of freedom for aligning metasurface resonances with molecular vibrational bands beyond the discrete geometries defined during fabrication.

Moreover, the thermal tuning process is reversible and free of measurable hysteresis. A representative cooling cycle for $\mathrm{MS_1}$ is shown in the inset of Fig. 3b, where the resonance progressively returns to its initial room-temperature position upon cooling. The extracted peak resonance positions for both heating and cooling cycles reveal nearly overlapping temperature dependence across all metasurfaces, further confirming negligible hysteresis and consistent thermo-optic response (see Supplementary Material Fig. S2). The reproducible recovery of the spectral response confirms that the thermo-optic modulation is reversible and that the metasurface structure remains optically stable during thermal cycling. Moreover, we did not observe any measurable structural degradation or delamination over repeated heating–cooling cycles, indicating robust mechanical integrity of the stencil-patterned Ge resonators on calcium fluoride.

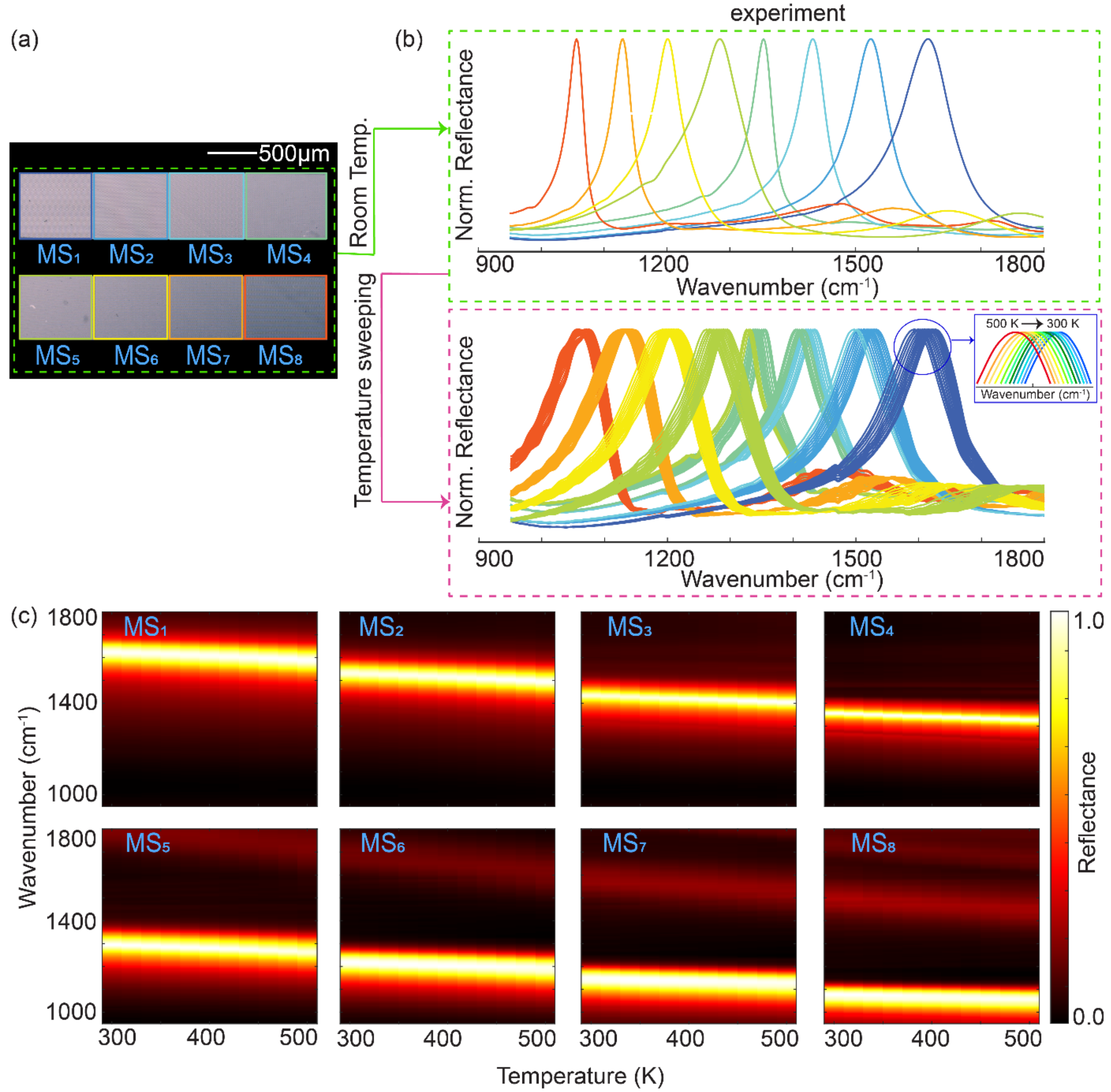

**Figure 3. Thermal tunability of the Ge pillar metasurface resonances for on-demand spectral reconfigurability. a,** Optical micrograph of the top two rows of a 4 × 3 metasurface array ($\mathrm{MS}_1 - \mathrm{MS}_8$). **b,** Top panel shows normalized reflectance spectra from $\mathrm{MS}_1 - \mathrm{MS}_8$ with varying unit cell geometry measured at room temperature, and the bottom panel shows the temperature-dependent reflectance spectra of the same metasurface array measured at a temperature range from 300 K to 500 K. The resonances exhibit continuous red shift with increasing temperature, driven by the thermo-optic response of Ge, exhibiting an average thermal sensitivity of ~0.18 $\mathrm{cm}^{-1}$/K (total ~36 $\mathrm{cm}^{-1}$ shift over 200 K). The inset illustrates a representative spectral blue shift upon cooling the metasurface, demonstrating tuning reversibility in both heating and cooling cycles. **c,** Normalized measured reflectance spectral colormaps showing the resonance evolution in $\mathrm{MS}_1 - \mathrm{MS}_8$ with temperature increase from 300 to 500K.

To visualize spectral evolution across the array, Fig. 3c presents two-dimensional experimental reflectance colormaps for $\mathrm{MS}_1 - \mathrm{MS}_8$ as functions of wavenumber and temperature. The bright yellow band in each panel indicates the dynamically tuned resonance peak, while darker regions represent off-resonance suppression. The yellow bands continuously redshift as temperature increases from 300K to 500K across all metasurfaces, indicating that the thermo-optic response is governed primarily by the intrinsic material properties of Ge rather than geometry-dependent effects. The dual spectral resonance control of the metasurfaces, both geometric tuning at the fabrication and dynamic reconfigurability post-fabrication achieves near-continuous spectral coverage across the fingerprint spectrum.

***Broadband, spectrally uniform SEIRAS with dynamic metasurface resonance sweeping***

To demonstrate SEIRAS using dynamically tunable and sharp Ge pillar array resonances, we investigated the interaction between the metasurface modes and the C–O–C stretching vibration of polymethyl methacrylate (PMMA) near 1148 $cm^{-1}$. PMMA serves as a model analyte owing to its well-characterized vibrational modes in the mid-IR and its relevance as a representative polymeric material. A thin PMMA layer (~70 nm) was spin-coated onto the metasurface array, forming a conformal film, suitable for label-free vibrational spectroscopy.

We first establish photonic resonance–molecular vibration alignment as the governing mechanism underlying SEIRAS (Fig. 4). By comparing reflectance spectra measured with and without the PMMA layer, we observe that both the magnitude and line shape of the extracted absorbance (EA) are dictated by the spectral overlap between the metasurface resonance and the molecular vibrational mode. Under on-resonance conditions, where the metasurface resonance aligns with the PMMA C–O–C vibration (~1148 $cm^{-1}$), the EA exhibits a maximum amplitude and a symmetric spectral profile, indicative of efficient resonance–vibration coupling. In contrast, off-resonance detuning leads to a ~25–30% reduction in EA amplitude and the emergence of asymmetric spectral features, reflecting diminished coupling strength originating from detuning. These observations highlight that precise spectral alignment is essential for achieving high-fidelity SEIRAS, because even a slight detuning degrades both enhancement efficiency and spectral accuracy.

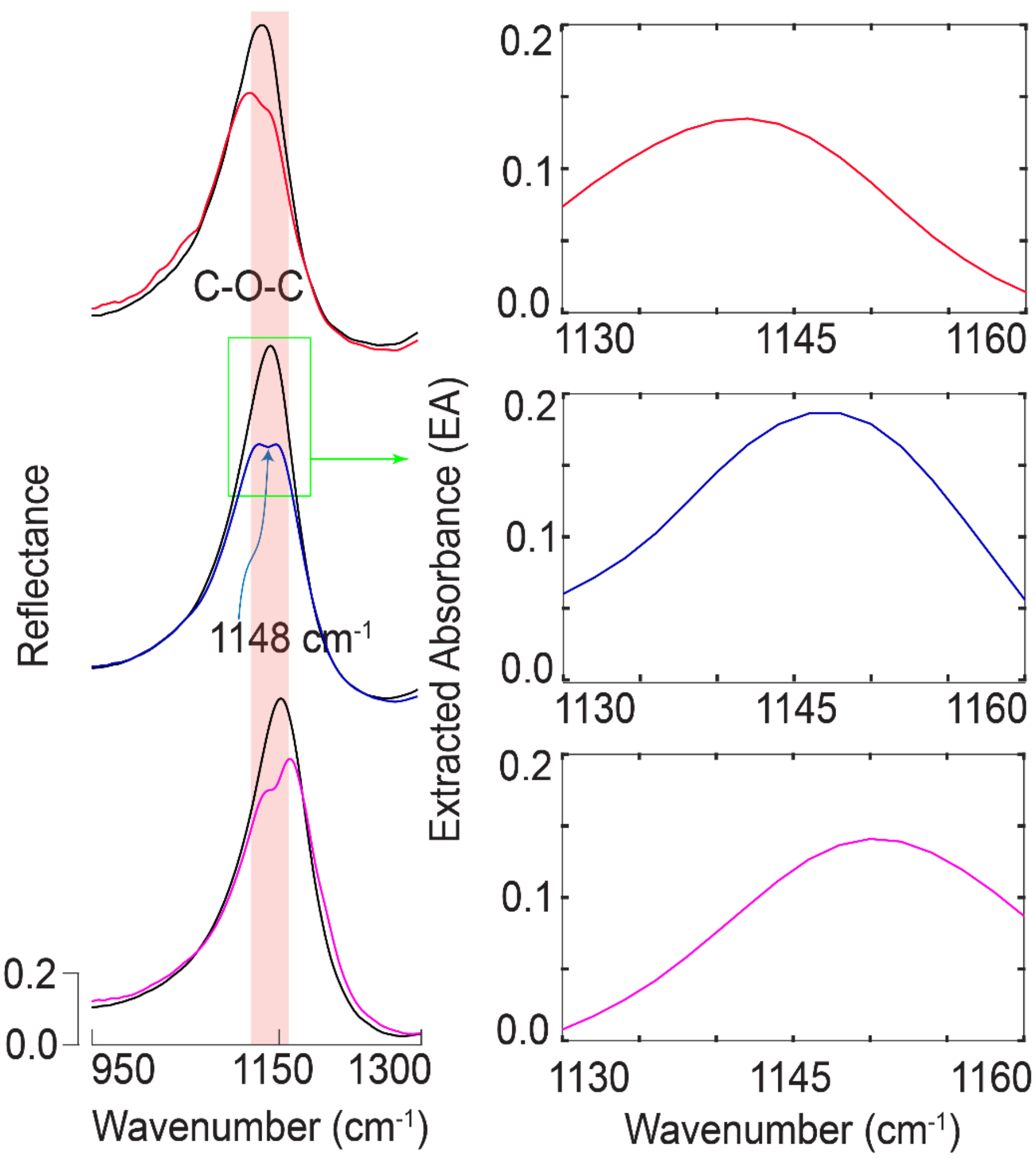

**Figure 4. Photonic resonance–molecular vibration alignment for efficient SEIRAS response. Left:** Experimental reflectance spectra of the Ge pillar array measured without (black) and with a ~70 nm PMMA overlayer (colored curves), shown for three representative detuning conditions. The shaded region highlights the PMMA C–O–C vibrational band (~1148 $cm^{-1}$). The top and bottom panels correspond to slightly off-resonant conditions, where the metasurface resonance is detuned from the molecular vibration, resulting in weaker and asymmetric reflectance modulation. The middle panel corresponds to on-resonant condition, where spectral overlap produces a pronounced reflectance dip. **Right:** Extracted absorbance (EA) obtained from differential reflectance of bare and PMMA coated metasurfaces for three detuning conditions. On-resonance (zero detuning) exhibits a symmetric line shape, whereas off-resonant cases (top and bottom) show reduced amplitude and asymmetric features, indicative of weaker resonance–vibration coupling.

We next leverage the dynamic and continuous controllability of our metasurface resonances to realize on-resonant SEIRAS across a broad spectral fingerprint region (Fig. 5), thereby, addressing a key limitation of static metasurface platforms. Figure 5a shows an infrared image of a metasurface array consisting of four distinct metasurfaces ($MS_{1\text{-}4}$ recorded at 1150 $cm^{-1}$), each supporting a slightly offset resonance at room temperature. Figure 5c presents the temperature-dependent reflectance spectra of the four metasurfaces measured without PMMA, demonstrating continuous spectral coverage spanning 1100–1215 cm$^{-1}$. After coating the array with a PMMA layer, we acquired a second set of temperature-dependent spectra from the same four metasurfaces under identical thermal conditions. Absorbance was then extracted by comparing spectra measured with and without PMMA, defined as the relative change in reflectance at the resonance peak for each temperature and metasurface configuration, thereby isolating the molecular absorption contribution as illustrated in Fig. 5b. As the thermally tuned resonance traverses the PMMA vibrational bands, the absorbance evolves accordingly, and we obtain a set of discrete spectral points spanning 1100–1215 cm$^{-1}$ (Fig. 5d). This enables reconstruction of PMMA's vibrational spectrum, which closely matches the reference profile and resolves key features near 1148 cm$^{-1}$ and 1192 cm$^{-1}$. The temperature-driven resonance sweep thus enables continuous spectral sampling and direct mapping of molecular absorption.

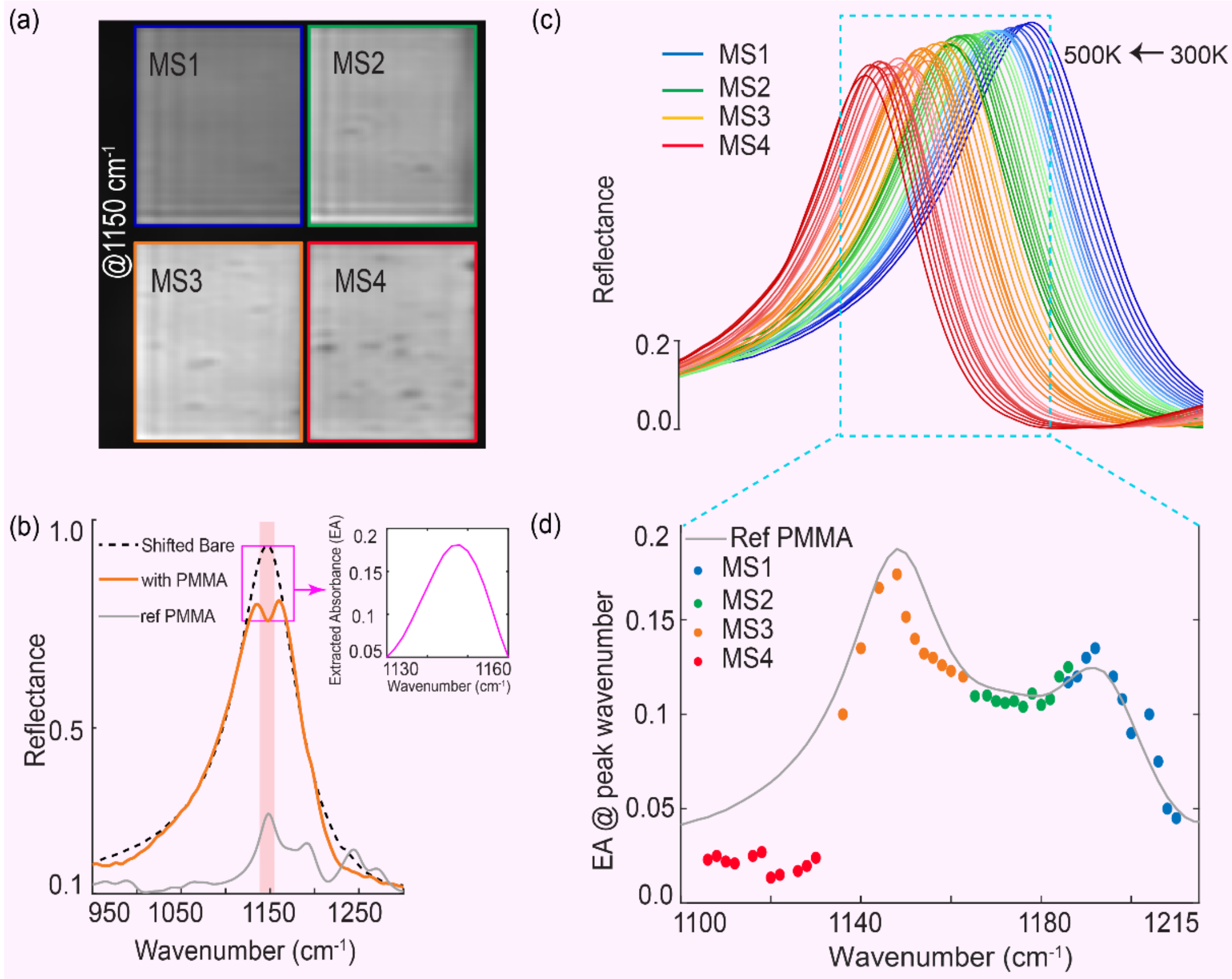

**Figure 5. SEIRAS of PMMA thin films using thermally tunable Ge pillar metasurfaces. a,** Mid-IR image showing a metasurface array of four, recorded at 1150 cm$^{-1}$. Colored metasurface frames correspond to their datasets shown in (b,c). **b,** Measured reflectance spectra from a single metasurface with and without 70-nm PMMA film coating. A distinct vibrational band centered at 1148 cm$^{-1}$, attributed to the C–O–C stretching mode of PMMA, couples to metasurface resonance and modulates its resonance peak. Inset shows extracted absorbance (EA) comparing before and after PMMA coating of the metasurface resonance from the pink shaded spectral window. Gray line shows PMMA's reference spectrum. **c,** Temperature-dependent reflectance response of the four metasurfaces at temperatures from 300 to 500 K, with each metasurface resonance sweeping a ~22 cm$^{-1}$ spectral range. **d,** PMMA's absorbance spectrum extracted as in (b) from temperature-tuned resonances from the four metasurfaces shown in (a). Each colored circular data point represents the absorbance at the resonance peak of the metasurfaces obtained from temperature-dependent measurements, extracted from reflectance spectra measured with and without a PMMA film, respectively. The continuous resonance sweeping enables to retrieve PMMA's vibrational fingerprints across the 1100 -1215 $cm^{-1}$ spectral range.

In summary, we developed a thermally reconfigurable dielectric metasurface platform based on all-dielectric Ge pillar arrays fabricated by scalable stencil lithography, enabling practical and efficient SEIRAS. By combining geometric and thermal tuning, we demonstrated continuous sweeping of sharp metasurface resonances across the molecular fingerprint region. We first established how the SEIRAS efficiency depends critically on precise spectral alignment between the photonic resonance and the molecular vibrational band. We then leveraged dynamic resonance control to achieve continuous on-resonant SEIRAS detecting PMMA's absorbance spectrum across the 1100–1215 $cm^{-1}$ region, containing its C-O-C stretch band.

Our approach addresses two major limitations of existing metasurfaces that are used for SEIRAS. First, we achieved scalable manufacturing of large area dielectric metasurface arrays using stencil lithography, where the shadow masks are mass manufactured using standard UV lithography. Second, our platform ensured uniformly strong light-matter interactions continuously across the broadband molecular fingerprint region via on-resonance SEIRAS. By integrating static geometric design with dynamic spectral tunability, our approach enables precise molecular fingerprinting in compact photonic architectures and establishes a practical route towards integrated, sensitive and high-resolution mid-infrared spectroscopy.

## Methods

### Numerical Simulations

Full-wave electromagnetic simulations were performed using the finite-element frequency-domain solver in CST Microwave Studio 2023 (Dassault Systèmes, France) to compute the transmittance spectra and Electric field distributions shown in Figs. 2b–d. To evaluate the scattering cross-section contributions from Cartesian multipoles, we employed Lumerical FDTD in combination with a custom MATLAB implementation of the multipole decomposition formalism[40]. All simulations were carried out using periodic boundary conditions in the x and y directions, with perfectly matched layers (PMLs) along the z-axis. The structure was excited using a normally incident plane wave with a Gaussian temporal envelope launched from the top. We used the temperature-dependent refractive index of Germanium to simulate the spectral response of the metasurface[39].

### Stencil mask fabrication

The gradient Ge pillar metasurfaces were fabricated using a stencil lithography process that enables resist-free pattern transfer onto infrared-transparent substrates. The stencil mask was fabricated on a silicon wafer to create a freestanding silicon nitride ($Si_3N_4$) membrane containing an array of circular apertures that define the metasurface geometry. The fabrication of the stencil mask begins with a 150 mm double-side-polished Si wafer (310 µm thick, ⟨100⟩ orientation), onto which a ~400 nm $Si_3N_4$ layer is deposited on both sides using low-pressure chemical vapor deposition (LPCVD). Standard photolithography (365 nm wavelength) is then used to pattern the aperture array on the front side of the wafer. The pattern is transferred into the $Si_3N_4$ layer through reactive ion etching (RIE), forming the circular openings that serve as the stencil apertures. Subsequently, the backside of the wafer is lithographically patterned to define the membrane windows, and the underlying silicon is removed using wet etching to release freestanding $Si_3N_4$ membranes. These suspended membranes act as mechanically robust stencil masks for subsequent deposition steps. The fabrication workflow follows the wafer-scale membrane release strategy previously reported for $Si_3N_4$ membrane devices[14].

**PMMA Spin Coating and Thickness Characterization**

A 70 nm-thick PMMA layer (PMMA 950 A2, Kayaku Advanced Material, Massachusetts, USA) was deposited using spin coating at 4000 rpm. To measure the PMMA thickness, each metasurface coated with PMMA was accompanied by a bare Si chip of the same size, which was also coated with PMMA under identical conditions. The thickness of the PMMA layer on the bare Si substrates was measured using an ellipsometer (J.A. Woollam, Nebraska, USA) and a reflectometer (Filmetrics, California, USA).

**Optical Characterization**

Mid-infrared optical characterization was performed in reflection geometry using a quantum cascade laser–based discrete frequency IR microscope (DRS Daylight Solutions, CA, USA) integrating four tunable sources to cover the 950–1800 $cm^{-1}$ range with 2 $cm^{-1}$ spectral resolution. During acquisition, the sample chamber was continuously purged with dry nitrogen to eliminate water vapor. The Ge pillar metasurface was illuminated with collimated, linearly polarized light at normal incidence. Spectral imaging was carried out using two different 12.5× infrared objectives: one with a numerical aperture (NA) of 0.7 and a field of view of 650 µm × 650 µm; and another with a 0.3 NA and a larger field of view of 2 mm × 2 mm. Both configurations employed an uncooled microbolometer focal plane array (480 × 480 pixels) for detection. The metasurface chip was mounted on a thermally controlled ceramic stage with programmable power input, enabling dynamic tuning from 300 K to 500 K. All measurements were background-referenced to Au.

**Acknowledgment**

F.Y. acknowledges financial support from the National Institutes of Health (grant no. R21EB034411, R35GM156309) and the National Science Foundation (grant no. 2401616). The authors gratefully acknowledge use of facilities and instrumentation in the UW-Madison Wisconsin Center for Nanoscale Technology. The Center (wcnt.wisc.edu) is partially supported by the Wisconsin Materials Research Science and Engineering Center (NSF DMR-2309000) and the University of Wisconsin-Madison.

**Data availability**

The data supporting the findings of this study are available from the corresponding author upon reasonable request.